\documentclass[runningheads]{llncs}

\usepackage{graphicx}

\newcommand{\etal}{\textit{et al}.}
\usepackage{subcaption}
\usepackage{amssymb}
\usepackage{pifont}
\usepackage{array}
\usepackage{hyperref}
\usepackage{wrapfig}
\usepackage{makecell}
\usepackage{multirow}
\newcolumntype{P}[1]{>{\centering\arraybackslash}p{#1}}
\usepackage{xcolor}

\usepackage{xspace}
\newcommand{\sysname}{ConGISATA}

\begin{document}

\title{\sysname: A Framework for Continuous Gamified Information Security Awareness Training and Assessment}

\titlerunning{\sysname: A Framework for Continuous Gamified ISA}


\author{Ofir Cohen \and
Ron Bitton \and
Asaf Shabtai \and
Rami Puzis}

\authorrunning{O. Cohen et al.}

\institute{
Software and Information Systems Engineering and Cyber@BGU,\\
Ben-Gurion University of the Negev, Beer-Sheva, Israel \\
\email{\{cofir,ronbit\}@post.bgu.ac.il}, 
\email{\{shabtaia,puzis\}@bgu.ac.il}}

\maketitle              

\newcommand\extrafootertext[1]{%
    \bgroup
    \renewcommand\thefootnote{\fnsymbol{footnote}}%
    \renewcommand\thefootnote{}%
    \footnotetext{#1}%
    \egroup
}

\extrafootertext{This version of the contribution has been accepted for publication, after peer review, but is not the Version of Record and does not reflect post-acceptance improvements, or any corrections. The Version of Record is available online at: \url{https://doi.org/10.1007/978-3-031-51479-1_22}. Use of this Accepted Version is subject to the publisher’s Accepted Manuscript terms of use \url{https://www.springernature.com/gp/open-research/policies/accepted-manuscript-terms}.
 \\
This paper received the \textbf{Distinguished Paper Award} at ESORICS 2023.}

\begin{abstract}
The incidence of cybersecurity attacks utilizing social engineering techniques has increased. 
Such attacks exploit the fact that in every secure system, there is at least one individual with the means to access sensitive information. 
Since it is easier to deceive a person than it is to bypass the defense mechanisms in place, these types of attacks have gained popularity.
This situation is exacerbated by the fact that people are more likely to take risks in their passive form, i.e., risks that arise due to the failure to perform an action. 
Passive risk has been identified as a significant threat to cybersecurity.
To address these threats, there is a need to strengthen individuals' information security awareness (ISA). 
Therefore, we developed \sysname\space - a continuous gamified ISA training and assessment framework based on embedded mobile sensors; a taxonomy for evaluating mobile users' security awareness served as the basis for the sensors' design. 
\sysname's continuous and gradual training process enables users to learn from their real-life mistakes and adapt their behavior accordingly.
\sysname\space aims to transform passive risk situations (as perceived by an individual) into active risk situations, as people tend to underestimate the potential impact of passive risks.
Our evaluation of the proposed framework demonstrates its ability to improve individuals' ISA, as assessed by the sensors and in simulations of common attack vectors.

\end{abstract}

\section{Introduction}
\label{chap:intro}
Defense mechanisms are deployed to prevent attackers from performing malicious activities such as hacking into networks, accessing sensitive information, and compromising computerized systems. 
Social engineering (SE) refers to techniques aimed at manipulating people into performing actions that help an attacker bypass state-of-the-art defense mechanisms~\cite {kumar2015social}.
The ease with which the human factor can be exploited has resulted in numerous cyberattacks caused by human error~\cite{kelly2017almost,selvam2020human}. 
For mobile users, SE is one of the main attack vectors~\cite{Threat-Report-2023}, and given the prevalence of smartphones today, SE poses a significant threat to society.

Approaches for mitigating the risk posed by cybersecurity attacks utilizing SE techniques consist of two essential components: assessing information security awareness (ISA) and improving it.

Various methods can be used to assess ISA, the most common being questionnaires~\cite{hart2020riskio,chapman2014picoctf,jaffray2021sherlocked}. 
However, questionnaires require users' active involvement and collaboration; moreover, they are subjective and prone to bias, as they rely on self-reported behavior~\cite{redmiles2018asking}. 
Despite their widespread use, questionnaires have been shown to be an unreliable measurement tool for ISA~\cite{bitton2020evaluating}.

Challenges involving simulations of common attacks are also used to measure ISA. 
The primary advantage of this type of assessment is that it measures users' ability to handle real-life attack scenarios. 
However, challenges also have a major limitation:
they do not consider users' context (e.g., opening an email from home versus opening an email at work).
Since human behavior often depends on the context, these methods are inherently less accurate~\cite{solomon2022contextual}. 

To address the limitations of existing ISA assessment methods, Bitton~\etal~\cite{bitton2018taxonomy,bitton2020evaluating} proposed a taxonomy for mobile users' security awareness that defines a set of measurable criteria organized by technological focus areas. These criteria are measured by a mobile agent that collects data from sensors on the users' devices. 
The sensors are mapped to the taxonomy's criteria, and a final passive ISA score is produced by aggregating their outputs. 
This ISA score can be changed dynamically based on continuous sensor readings.
In this research, we use this sensor-based approach, along with challenges associated with three common attack vectors, to assess ISA.

Typically, ISA is improved by participating in security awareness programs (workshops) or performing challenges with feedback.
However, the previously mentioned limitation also applies when challenges are used to improve ISA.
In many cases, efforts aimed at improving ISA evoke fear, which has been shown to be counterproductive at times; such efforts also result in `security fatigue,' in which people tire of being presented with security procedures and processes~\cite{bada2019cyber}.

Gamification is a technique often used to overcome the limitations described above. 
Deterding~\etal~\cite{deterding2011game} defined gamification as ``the use of game design elements in non-game contexts,"  and Hamari~\etal~\cite{hamari2014does} reviewed many gamification studies and concluded that this method works well in various fields, particularly for improving learning and training sessions.
As a result, the use of gamification to increase ISA has grown, leading to the development of various gamified solutions for this purpose~\cite{alqahtani2020design,luh2020penquest,yasin2018improving}.

Nevertheless, standard gamification alone is insufficient.
A literature review performed by Böckle~\etal~\cite{bockle2017towards} highlighted the problem of the ``one size fits all" approach, which may result in declining engagement and loss of interest in overly simple challenges.
To overcome this, the authors suggested using adaptive gamification that dynamically re-engages users.
Our approach for improving ISA utilizes adaptive gamification through a personalized feedback loop tailored to the outputs of the sensors.

A behavioral aspect that was not considered in previously proposed gamified solutions is passive/active risk-taking.
Studies have classified risks as either active or passive risks~\cite{keinan2012leaving,keinan2017perceptions}.
Active risk describes actions people take that put them at risk, while passive risk is ``\textit{risk brought on or
magnified by inaction or avoidance}." 
One example from the cybersecurity domain is the risk of having malware on your mobile device. 
In its active form, this risk derives from the possibility of unknowingly downloading a malicious file, whereas in its passive form, it stems from failing to install anti-malware software on the device in advance.
These studies showed that passive risks are perceived as being less risky than equivalent active risks. Therefore, our framework aims to reduce passive risk-taking (PRT), by transforming passive risks into active risks. 
By deducting game points from users who fail in a passive-risk-related scenario, we impose an immediate punishment on passive behaviors.
By doing so, we can help users gradually overcome the human tendency to overlook passive risks.

In this research, we propose \sysname, a continuous gamified ISA training and assessment framework, which addresses the problems of existing gamification-based methods described above. 
Our approach is implemented by a mobile agent (an app) that collects data from the set of sensors used in the taxonomy and assessment method of Bitton~\etal~\cite{bitton2018taxonomy,bitton2020evaluating}. 
The app has a graphical user interface with the key components of a gamified system: a detailed home screen, a leaderboard, and a learning screen. 
The learning screen is composed of sections, one for each criterion in the taxonomy. 
For each criterion, there is a score and a link to an article or blog post that should help users improve their behavior with regard to the criterion. 
The scores on this screen are updated daily according to the sensors' readings and highlight the areas in which the user needs to improve. 
Challenges are also presented throughout the learning process to help assess users' ISA as they progress.

To evaluate the proposed framework, we performed an extensive experiment involving 70 subjects, each of whom installed our mobile app on their smartphone and used \sysname\space for a period of five weeks.
We compared our method with a baseline method inspired by methods commonly used in academia and industry today. 
In the baseline method, users were provided personalized articles/blog posts based on their performance in the challenges, without taking the sensor data into account. 
Our results show that users who were trained using the \sysname\space framework had greater improvement than those trained using the baseline method for almost all criteria of the ISA taxonomy. 
In addition, a significant correlation between the use of our app and users' ISA improvement was observed.
Importantly, by using simulations of three common attack vectors, we found that \sysname\space helps users deal with real-life SE scenarios.

Our contributions can be summarized as follows:(1) We propose a novel framework for improving and assessing mobile users' ISA. (2) To the best of our knowledge, we are the first to take continuous sensor readings and show their impact on improving ISA in an adaptive gamification setting. (3) We empirically demonstrate the importance of considering passive risk-taking in the ISA training domain.

\section{Background}
\label{sec:background}
\noindent\emph{Active versus passive risk-taking:}
Keinan~\etal~\cite{keinan2012leaving} established passive risk as a unique and separate construct.
The authors provide the following explanation:``\textit{People are often held less responsible for their omissions than for their commissions. 
This lack of perceived responsibility may lower the motivation to act. 
People are usually less likely to do something if they believe they will not be held accountable for failing to do it. 
However, risk aversion often increases with personal accountability, since accountability stimulates self-critical forms of thought and increases awareness of one's own judgment processes. 
It seems plausible that once people feel accountable they process information better, realize that they are in a risky situation, and be motivated to act to avoid risk.}"

A follow-up paper~\cite{keinan2017perceptions} showed that a passive risk is judged as less risky than a completely equivalent active risk. 
For example, the following scenario was presented in both active and passive forms: actively parking your car in a restricted zone or not moving your car once you realize it is parked in a restricted zone.
When asked to rate scenarios by risk level, in its active form this scenario was rated as riskier than in its passive form.

The authors suggest that ``\textit{this inferior ability to devote attention to the absence of events leads to passive risks being less available to our consciousness, to be underestimated, and thus to be perceived as less risky. 
We need to be motivated to devote attention to passive risks.}"
The authors add the following recommendation: ``\textit{The findings of the current research suggest stressing people's personal responsibility for complying with recommended preventive measures may raise risk perception and increase preventive action.}"

Finally, Arend~\etal~\cite{arend2020passive} examined how self-reported passive risk behavior predicts cybersecurity behavioral intentions and their relation to actual cybersecurity behavior.
This series of three studies showed that passive risk had a notable impact on cybersecurity intentions, meaning that high passive risk scores were associated with low adherence to safe cybersecurity behavior.
It was also shown that behavioral choices related to cybersecurity are highly correlated with a tendency to take passive risks.
Overall, these studies established that passive risk tendencies are an important factor in the context of cyber behavior.

\section{Related Work}
Every gamified approach for mitigating the risk posed by SE attacks consists of two essential components: measuring ISA and improving it. 

Questionnaires are the most common means of measuring ISA, with the vast majority of prior studies on this topic relying on them~\cite{gjertsen2017gamification,newbould2009playing,denning2013control,dincelli2020choose,scholefield2019gamification,omar2021malware,wu2021assessing}.
Despite their widespread use, they tend to be an unreliable measurement tool for behavior because of their subjective nature. Additionally, they are prone to bias, as they rely solely on self-reported behavior~\cite{redmiles2018asking}.

A more advanced method of evaluating ISA is to use attack simulations (also referred to as challenges).
Despite their inability to consider users' context, the employment of challenges to assess ISA during security awareness training is extremely valuable, as it provides important insights into authentic user behavior. 
Their application in the literature, however, has been limited~\cite{canham2022phish}.
Our framework utilizes three different types of challenges:  phishing, permission attacks, and impersonation. 
When using our framework, users do not know when or how they are presented with these challenges; this ensures that the challenges are as natural and objective as possible.
Additionally, the framework collects data from sensors in users' everyday environments to examine aspects of their ISA in real-life settings, outside of controlled laboratory conditions.

When it comes to improving ISA using gamification, two core elements distinguish current gamified solutions: training duration and personalization of the content.
Most of the gamified training mentioned in the literature was performed for a single brief session and utilized a physical board/card game, which is a difficult requirement for long-term training (over a period of weeks)~\cite{newbould2009playing,denning2013control}. 
We only identified one paper with a longer training process -- that of Alahmari~\etal, where the training took place for two weeks~\cite{alahmari2022moving}. 
Our framework is designed to achieve long-term behavioral change through continuous learning over several weeks or months, without requiring physical attendance at training sessions.

Böckle~\etal~\cite{bockle2017towards} highlighted the problem of the ``one size fits all" approach in gamified solutions for improving ISA and suggested the use of personalization.
Heid~\etal~\cite{heid2020raising} created a gamified prototype that poses questions related to security and privacy issues associated with apps installed on the user's smartphone. 
A quiz engine providing multiple choice questions regarding known vulnerabilities and app properties was implemented using Appicaptor, a mobile application analysis platform that performs static and dynamic app tests. 
The question engine automatically generates questions from Appicaptor's database content, which are personalized for the users based on the apps installed on their smartphones. 
However, this work is limited, because only one sensor served as a source of information, the proposed method was only tested within the research group, and it relies on an external data source that is not publicly available, preventing its reproducibility. 

Our literature review failed to identify any other papers utilizing personalization besides the work of Heid~\etal{} mentioned above. 
Our framework generates scores for each user based on their weaknesses, as measured using multiple sensors. 
We evaluated the framework's impact in a comprehensive experiment spanning several weeks. 
All the materials used are publicly available and presented in the appendix.
Furthermore, our gamified solution is the only method that demonstrates how passive risks can be transformed into active ones, which is a key contribution of our research.

A summary of the related work is provided in Table~\ref{table:related_work}.
 
\begin{table}
\caption{Summary of related work}
\label{table:related_work}
\scriptsize
\centering
\begin{tabular}{|l|l|P{0.6cm}|P{0.6cm}|P{0.6cm}|c|c|c|}
\hline
 \textbf{Paper} & \textbf{Platform} & 
 \rotatebox[origin=c]{90}{\parbox{2.1cm}{\textbf{Personalization}}} & 
 \rotatebox[origin=c]{90}{\parbox{2.1cm}{\textbf{Considers \\ PRT}}} & 
 \rotatebox[origin=c]{90}{\parbox{2.1cm}{\textbf{Continuous \\ Learning }}} & 
 \rotatebox[origin=c]{90}{\parbox{2.1cm}{\textbf{Questionnaires}}} & 
 \rotatebox[origin=c]{90}{\parbox{2.1cm}{\textbf{Attack \\ Simulations}} }& 
 \rotatebox[origin=c]{90}{\parbox{2.1cm}{\textbf{Sensors}}} \\ \hline

Newbould \etal~\cite{newbould2009playing}, 2009 & Board game & \ding{55} & \ding{55} & \ding{55} & \checkmark & \ding{55} & \ding{55}\\ \hline
Denning \etal~\cite{denning2013control}, 2013 & Tabletop card game & \ding{55} & \ding{55} & \ding{55} & \checkmark & \ding{55} & \ding{55}\\ \hline
Gjertsen \etal~\cite{gjertsen2017gamification}, 2017 & Exercises & \ding{55} & \ding{55} & \ding{55} & \checkmark & \ding{55} & \ding{55}\\ \hline
Scholefield \etal~\cite{scholefield2019gamification}, 2019 & Mobile (Android) game & \ding{55} & \ding{55} & \ding{55} & \checkmark & \ding{55} & \ding{55}\\ \hline
Dincelli \etal~\cite{dincelli2020choose}, 2020 & Interactive storytelling & \ding{55} & \ding{55} & \ding{55} & \checkmark & \ding{55} & \ding{55}\\ \hline
Heid \etal~\cite{heid2020raising}, 2020 & Multiple choice quizzes & \checkmark & \ding{55} & \checkmark & \ding{55} & \ding{55} & \checkmark\\ \hline
Omar \etal~\cite{omar2021malware}, 2021 & Educational quizzes & \ding{55} & \ding{55} & \ding{55} & \checkmark & \ding{55} & \ding{55}\\ \hline
Wu \etal~\cite{wu2021assessing}, 2021 & Multiple choice quizzes & \ding{55} & \ding{55} & \ding{55} & \checkmark & \ding{55} & \ding{55}\\ \hline
Alahmari \etal~\cite{alahmari2022moving}, 2022 & Mobile app & \ding{55} & \ding{55} & \checkmark & \checkmark & \ding{55} & \ding{55}\\ \hline
Canham \etal~\cite{canham2022phish}, 2022 & Phishing simulations & \ding{55} & \ding{55} & \ding{55} & \checkmark & \checkmark & \ding{55}\\ \hline
Our method, 2023 & Mobile (Android) game & \checkmark & \checkmark & \checkmark & \ding{55} & \checkmark & \checkmark\\ \hline
\end{tabular}
\end{table}

\section{Proposed Method}
\label{chap:methodology}
\begin{figure}
    \centering
    \begin{subfigure}[t]{0.59\textwidth}
        \includegraphics[width=1\textwidth, keepaspectratio]{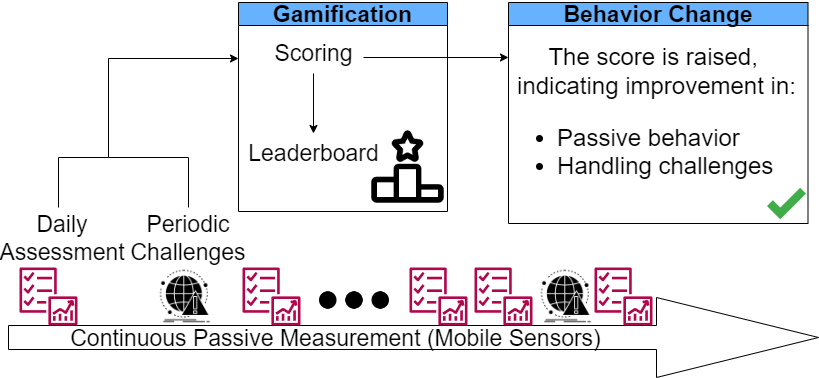}
        \caption{\sysname\space overview}
        \label{fig:overview}
    \end{subfigure}
    \begin{subfigure}[t]{0.39\textwidth}
        \includegraphics[width=1\textwidth, keepaspectratio]{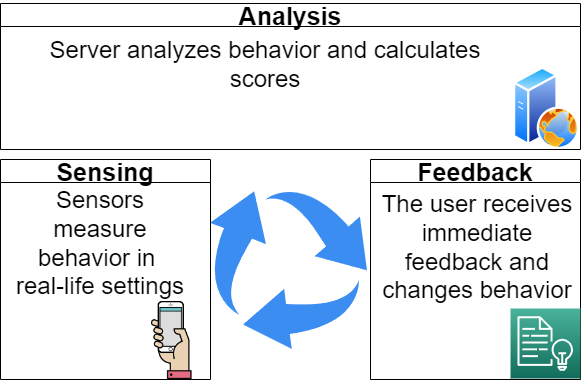}
        \caption{Daily feedback loop}
        \label{fig:feedbackloop}
    \end{subfigure}
    \caption{The \sysname\space security awareness training and assessment framework}
\end{figure}

In this section, we present \sysname. 
First we provide a high-level description of the framework (illustrated in Figure~\ref{fig:overview}), and then we elaborate on each component. 

In the framework, the following steps are performed in a process aimed at raising ISA:

  \noindent\emph{Calibration Period:} For each user, the game starts with a calibration period in which the user's initial security awareness score is assessed for each criterion in the taxonomy. 
  This assessment does not require the user to interact with the game, as it is performed using the mobile sensors and challenges described in Section~\ref{sec:assessing_isa}.
  Following this evaluation, the initial overall ISA score is presented to the user on the game's home screen, and a score for each criterion is presented on the learning screen.
  
  \noindent\emph{Training}: In this step, the user starts to interact with the framework in an attempt to gradually improve their behavior and raise their ISA scores. 
  Sensor measurement and challenges still occur in the background, resulting in daily changes to the user's ISA scores, which are presented to them. Training is performed cyclically, through a daily feedback loop, as follows: 
  
  \noindent\emph{Sensing} - Each day, the different aspects of the users' behavior are measured by obtaining the sensor values.

  \noindent\emph{Updating Scores} - The application's learning screen is always accessible and displays the users' scores for each criterion in the ISA taxonomy, along with their overall ISA score. 
  At midnight, these scores are updated to reflect the previous day's performance. 
  The learning screen also presents the score delta for each criterion, which is the change in the score between two consecutive days. 
  The score deltas enable the user to identify specific behavioral weaknesses (criteria with a negative score delta) and take corrective action.

  \noindent\emph{Articles and Blog Posts} - When the user is faced with a low score or a negative score delta for a criterion, they can obtain additional information about that specific focus area via the learning screen, which provides a link to an external, predetermined, and comprehensive article or blog post (for convenience, we refer to them as articles in the rest of the paper) on that subject.

  \noindent\emph{Behavior Change} - Upon reading the articles, the user will modify their behavior accordingly, improving their score over time, and climb the leaderboard.
  
  Figure~\ref{fig:feedbackloop} illustrates the daily feedback loop of passive ISA.  

\subsection{Assessing Mobile Users' ISA}
\label{sec:assessing_isa}
To generate an overall ISA score for each user, we measure two aspects of their behavior:
active and passive.
The active side refers to the user's ability to handle situations in which immediate action is required, as when facing an attack. Our framework measures this aspect using SE challenges.
The passive side refers to ongoing elements of the user's behavior that do not result in an immediate punishment if not performed, such as using a lock screen or deleting unused apps to avoid malware.
In our framework, we adapt the method proposed by Bitton~\etal ~\cite{bitton2018taxonomy,bitton2020evaluating} to generate a passive ISA score.
Instead, we generate an overall ISA score, which reflects both aspects, active and passive, as follows:

\noindent\emph{Assessing ISA Using Attack Simulations (Challenges):}
\label{subsec:challenges}
Each user is regularly presented with challenges derived from three attack vectors. 
These challenges assess the user's ability to handle real-life attack scenarios and ensure that this capability is also reflected in their overall ISA score. 
The challenges are presented in a randomized manner (in terms of both time and order) throughout the training process, to prevent detectable patterns.
The active score denotes the user's performance on SE challenges and is based on a scale of zero to 100. 
The score is derived from a moving window of the last $X$ challenges, where $X$ is determined based on the training duration. 
Each challenge is individually scored between zero, assigned for a failure to make the correct decision, and $100/X$, assigned for successful decision making. For example, for $X=5$, each challenge can contribute at most $100/5=20$ points to the overall active score.
Some challenges may involve two decision points, in which case the score assigned is $(100/X)/2$ if the user makes the correct decision at just one of the decision points. For example, a phishing challenge may include two decision points; the first is clicking on the unknown link to enter the phishing website, and the second is providing sensitive details such as login credentials. In such a case, if a user only clicks on the unknown link but does not provide any details, $(100/5)/2 = 10$ points will be added to the overall active score.

\noindent\emph{Assessing ISA Using Sensor Measurements:}
\label{subsec:isa_passive_asses}
Bitton~\etal~\cite{bitton2018taxonomy} developed a taxonomy to measure mobile users' ISA that classifies criteria by technological focus areas and psychological dimensions. 
Each focus area is further divided into sub-focus areas, and each of these sub-focus areas encompasses several security topics. 
For instance, the ``Applications" focus area is bifurcated into ``Application Installation" and ``Application Handling" sub-focus areas, with ``Untrusted Sources" as a security topic under ``Application Installation." 
The intersection of this security topic with the ``Confronting Behavior" psychological dimension leads to a specific criterion: ``\textit{Installs applications solely from trusted sources}."

Bitton~\etal~\cite{bitton2020evaluating} also proposed a framework for evaluating ISA, which employs a mobile agent with embedded sensors, a network traffic monitor, and cybersecurity challenges. 
Their framework, which is based on the ISA taxonomy, enables the computation of ISA scores at any given point. 
The study found that there was a difference between users' self-reported behavior and their actual behavior, highlighting the significance of monitoring real-life user behavior instead of relying solely on questionnaires.

In order to gain a deeper understanding of users' behavior in real-life scenarios, we included sensors based on the ISA taxonomy in our mobile application. 
These sensors periodically perform a thorough scan of users' devices and actions. 
By analyzing the resulting data, we can compute a user's passive ISA score and an individual score for each criterion, and identify their potential weaknesses. 
This knowledge allows us to provide the user with personalized feedback about their ISA scores and offer guidance on how to improve their security practices.

In the paper, we describe \sysname's use in training a group of users, which we believe is the more common scenario. 
The framework can be easily adapted to train individuals, however we do not discuss that in the paper.
The process of computing the passive score begins with a calibration period, during which each user's initial passive ISA score is obtained, without any prior training. 
After this period, the mean and standard deviation of the entire user group are calculated for each criterion in the taxonomy. 
During training, a new z-score (standard score) is computed daily for each user for each criterion, using the mean and standard deviation derived in the calibration period. 
The new z-scores are then averaged for each of the taxonomy's focus areas and are subsequently averaged again to obtain a final passive score for each user. 
Since the z-score is not meaningful to users, the cumulative probability function of the normal distribution is used to transform the z-score to a 0-100 scale.

\noindent\emph{Computing the Overall Score:}
The overall ISA score is the average of the active and passive scores.

\subsection{Gamification}
\label{sec:gamification}

To increase user engagement and optimize the effectiveness of training, we have incorporated essential gamification elements into our framework. Table~\ref{table:game} lists some of these key elements, along with their rationale, and explains how they have been implemented in \sysname.

\begin{table}
\caption{\sysname's gamification elements}
\label{table:game}
\scriptsize
\centering
\begin{tabular}{|P{1.6cm}|p{10.4cm}|}
\hline
\textbf{Element} &
 \textbf{Explanation} \\ \hline
Continuous Learning & Dunlosky~\etal~\cite{dunlosky2013improving} provided a comprehensive review of study techniques and assessed their effectiveness. 
One of the techniques covered is continuous learning, which was termed \textit{distributed practice} and defined as ``\textit{implementing a schedule of practice that spreads out study activities over time}."
Based on prior research, distributed practice was one of just two techniques to be rated by the authors as having high utility. 
It was assessed that distributed practice ``\textit{works across students of different ages, with a wide variety of materials, on the majority of standard laboratory measures, and over long delays}."
Focusing on the cybersecurity domain, the findings of Kumaraguru~\etal~\cite{kumaraguru2009school} align with those of Dunlosky~\etal, demonstrating the benefits of extended security training over condensed single sessions.
Based on these findings, we designed our game as a continuous learning process, unlike the common approach found in the literature of a single-session game. \\ \hline
Considers PRT & Following the research presented in Section~\ref{sec:background}, we implemented a penalty mechanism to discourage PRT, whereby users that fail to take preventive measures will face penalties, resulting in point deductions. This approach transforms PRT into active risk-taking, where users are held accountable for their inaction shortly after it occurs, regardless of whether or not any damage was incurred.
For instance, if our sensors detect that certain users have not installed anti-malware software, they will have points deducted, even if no malware has exploited this vulnerability on their devices. 
Furthermore, users will continue to face daily penalties until they address and fix the issue by installing anti-malware software, further discouraging avoidance behavior. \\ \hline
Levels / Progression & It is crucial to provide players with a clear indication that they are acquiring knowledge and advancing through the training process. 
We achieve this through a ranking system comprised of two elements: points and levels. Players earn points (reflected in their ISA score) by exhibiting good security practices, and as they accumulate more points, they move up to higher levels. Our framework assigns users to one of three levels based on their ISA score: ``beginner," ``intermediate," and ``pro." These levels do not change the difficulty of training and are only used to give the users the feeling that they are advancing. \\ \hline
Competition & Competition is a fundamental aspect of nearly every game, in contexts including security. 
Healthy competition can significantly enhance engagement and enjoyment among players and encourage individuals to surpass their previous performance. 
Our game incorporates competition through (1) a leaderboard that ranks players by points, providing insight into their standing relative to others; and (2) the points and levels mentioned above, promoting competition among players. \\ \hline
Adaptive Gamification Through Personalized Feedback / Guidance & Immediate personalized feedback is important to prevent player confusion and maintain their engagement in the game. 
Further guidance helps players progress and improve as the game continues. 
Immediate feedback in our game is in the form of the learning screen. 
Each event that causes points to be earned or deducted, such as a sensor discovering poor application handling behavior, is presented on the learning screen on the day on which the event occurred. 
Additional guidance is possible through a dedicated article on the event's topic. 
In addition, each user's scores appear on their learning screen, highlighting the areas requiring improvement. \\ \hline
Conciseness & The game's exercises should be brief and not take much of the players' time. 
In our game, the feedback is succinct and highlights the topics pertinent to each player. 
This approach reduces the time commitment for players and avoids redundant review of familiar material. \\ \hline
\end{tabular}
\end{table}

\section{Evaluation}
\label{chap:evaluation}
To evaluate the proposed framework, we performed a long-term experiment involving 70 undergraduate and graduate students who use their smartphones regularly. 
The subjects' ages ranged from 21 to 31, with a mean age of 25 and a median age of 26. 
The experiment involved the collection of sensitive personal information from subjects for a long period of time, including their browsing patterns. 
We took measures to preserve the subjects’ privacy and reduce any privacy risks associated with participating in the experiment. 
The experiment was approved by the Institutional Review Board (IRB), provided that:
(1) The subjects participated in the experiment, freely, at their own will.
The subjects received course credit in exchange for their participation. 
The subjects were fully aware of the type of data that would be collected and were allowed to withdraw from the study at any time.
(2) The data was encrypted before being transmitted between the server and the mobile app.
(3) The server was within the university domain, with restricted access and organizational defenses. 
(4) When possible, the sensitive data itself was not transmitted to the server (such as SMS contents), only the meta-data was (such as the number of SMS messages containing URLs).
During the experiment, we measured the subjects' behavior while operating their smartphones and exposed them to three types of SE attacks in 15 attack simulations. 
We then compared each subject's initial and final ISA scores, measuring the improvement achieved. We also examined how the participants' performance in responding to the challenges evolved during the training process. This section provides a detailed description of the evaluation process and results.

\subsection{Mobile Sensors}
To evaluate the passive aspects of subjects' behavior, we implemented multiple sensors using Android APIs and used them to assess various criteria from the taxonomy of Bitton~\etal{}. 
We did not assess all of the criteria for reasons of simplicity and privacy.
The criteria and the way they were assessed are presented in Table~\ref{table:criteria}. 
In some cases, we found that the corresponding sensor did not work well for a large number of subjects or the sensor had no influence on the score; for example, for criterion OS2, we found that all of our subjects did not root their device before or during the experiment. 
In such cases, we omitted these sensors and the criteria that correspond to them, and they are not included in our analysis of the results. 
In addition, 10 out of the 70 subjects either had a technical problem with their smartphone which prevented them from participating, did not use the app, or decided to withdraw from the study. 
These subjects were omitted from the results analysis as well. 
Finally, while a higher z-score usually indicates better performance, some of the criteria represent bad behaviors (such as criterion AI1). In such cases, indicated in Table~\ref{table:criteria} by having ``(lower is better)" in their means of assessment, we multiplied their z-score by -1, changing positive numbers into negative progression indicators.

\begin{table}
\caption{List of criteria assessed for the experiment}
\label{table:criteria}
\scriptsize
\centering
\begin{tabular}{|P{3cm}|p{8.8cm}|}
\hline
\textbf{Criterion} &
\textbf{Means of assessment} 
\\ \hline
AI1: Downloads apps from trusted sources & An app was considered trusted if it was downloaded from an official app store (such as Google Play). 
The score for this criterion is the number of untrusted apps found on the subject's device (lower is better). \\ \hline
AI2: Does not install apps that require dangerous permissions & The score for this criterion is the number of apps on the subject's device which require dangerous permissions, as classified by Android (lower is better). \\ \hline
AI3: Does not install apps with a low rating & We considered a low rating to be less than three and a half stars (out of five) in the Google Play store. The score for this criterion is the number of apps with a low rating found on the subject's device (lower is better). \\ \hline
AH1: Regularly updates apps & Google Play features the last date on which an app was updated. The score for this criterion is the number of apps that are not up-to-date found on the subject's device (lower is better). \\ \hline
AH3: Properly manages running/installed apps & An app is considered unused if the subject did not use the app for more than two weeks. The score for this criterion is the number of unused apps found on the subject's device (lower is better). \\ \hline
B1: Does not enter malicious domains & A domain is considered malicious if Google's safebrowsing API has classified it as such. The score for this criterion is the number of malicious domains the subject has entered in the last seven days (lower is better). \\ \hline
VC1: Does not open messages received from unknown senders & We monitored two message inboxes for each subject - SMS and Gmail’s spam inbox. An SMS is considered to be from an unknown sender if the sender of the SMS is not in the subject’s contact list. The SMS score is the percentage of how many unknown SMSs the subject has opened in the last seven days. Likewise, the Gmail score is the percentage of emails classified as spam by Gmail that were opened in the last 30 days. The final score for this criterion is the average of the SMS and Gmail scores (lower is better). \\ \hline
VC2: Does not click on links received from unknown senders & We considered an event to be of the 'clicking on links received from unknown senders' type if the following three conditions were met: (1) the subject opened a message from an unknown sender, as defined in VC1, (2) the message that was opened contained a URL, and (3) we also identified a transition between the SMS/Gmail apps and the browser app (Google Chrome), suggesting the subject has clicked on that URL. The score for this criterion is the number of times a subject has clicked on URLs from unknown senders in the last seven days (lower is better). \\ \hline
A2: Uses two-factor authentication mechanisms & A subject was considered to be using two-factor mechanisms if either a two-factor authentication app or an SMS (from the last seven days) indicating two-factor use was found on their device. The score for this criterion is one if the subject uses two-factor mechanisms and otherwise zero (higher is better). \\ \hline
A3: Uses password management services & The subject was considered to be using password management services if a password-managing app was found on their device. The score for this criterion is one if the subject uses password management services and otherwise zero (higher is better). \\ \hline
OS2: Does not root or jailbreak the device & We used a dedicated Android package (rootBeer) that implements various heuristics to determine whether or not a device is rooted. The score for this criterion is one if the subject has not rooted the device and otherwise zero (higher is better). \\ \hline
SS2: Uses anti-virus application regularly to scan the device & The score for this criterion is one if the subject has an anti-virus app installed on the device and otherwise zero (higher is better). \\ \hline
SS5: Uses PIN code, pattern, or fingerprint & A device was considered secured if a lock-screen was enabled. The score for this criterion is one if the subject's device is secured and otherwise zero (higher is better). \\ \hline
N1: Does not connect to unencrypted networks & A network was considered encrypted if a security protocol was enabled (such as WPS, WPA2). The score for this criterion is the number of unencrypted networks the subject has connected to in the last seven days (lower is better). \\ \hline
N3: Uses VPN services on public networks & The subject was considered to be using VPN services if a VPN app was found on their device. The score for this criterion is one if the subject uses VPN services and otherwise zero (higher is better). \\ \hline
PC1: Disables connectivity when not in use & The score for this criterion is the number of times in the last seven days that a connectivity channel (i.e., Bluetooth, Wi-Fi, NFC) was enabled for more than five minutes, without being connected (lower is better). \\ \hline
\end{tabular}
\end{table}

\subsection{Social Engineering Challenges}

\begin{figure}
    \centering
     \begin{subfigure}[t]{0.49\textwidth}
        \includegraphics[width=\textwidth, height=0.2\textheight]{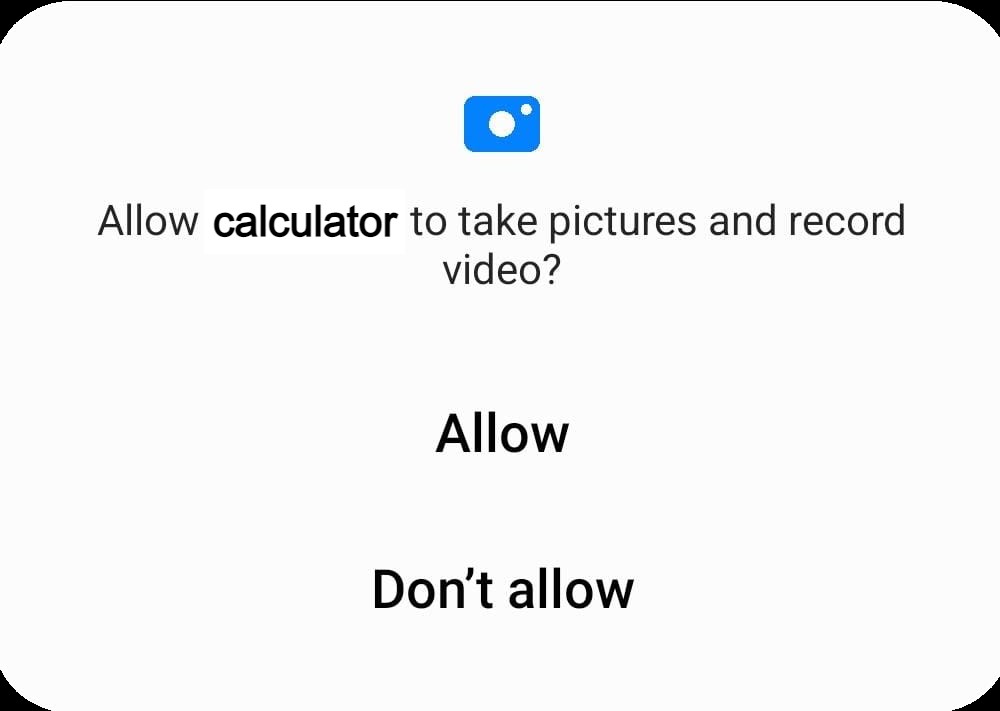}
        \caption{Permission attack}
        \label{fig:perm_atk}
    \end{subfigure}
    \begin{subfigure}[t]{0.49\textwidth}
        \includegraphics[width=\textwidth, height=0.2\textheight]{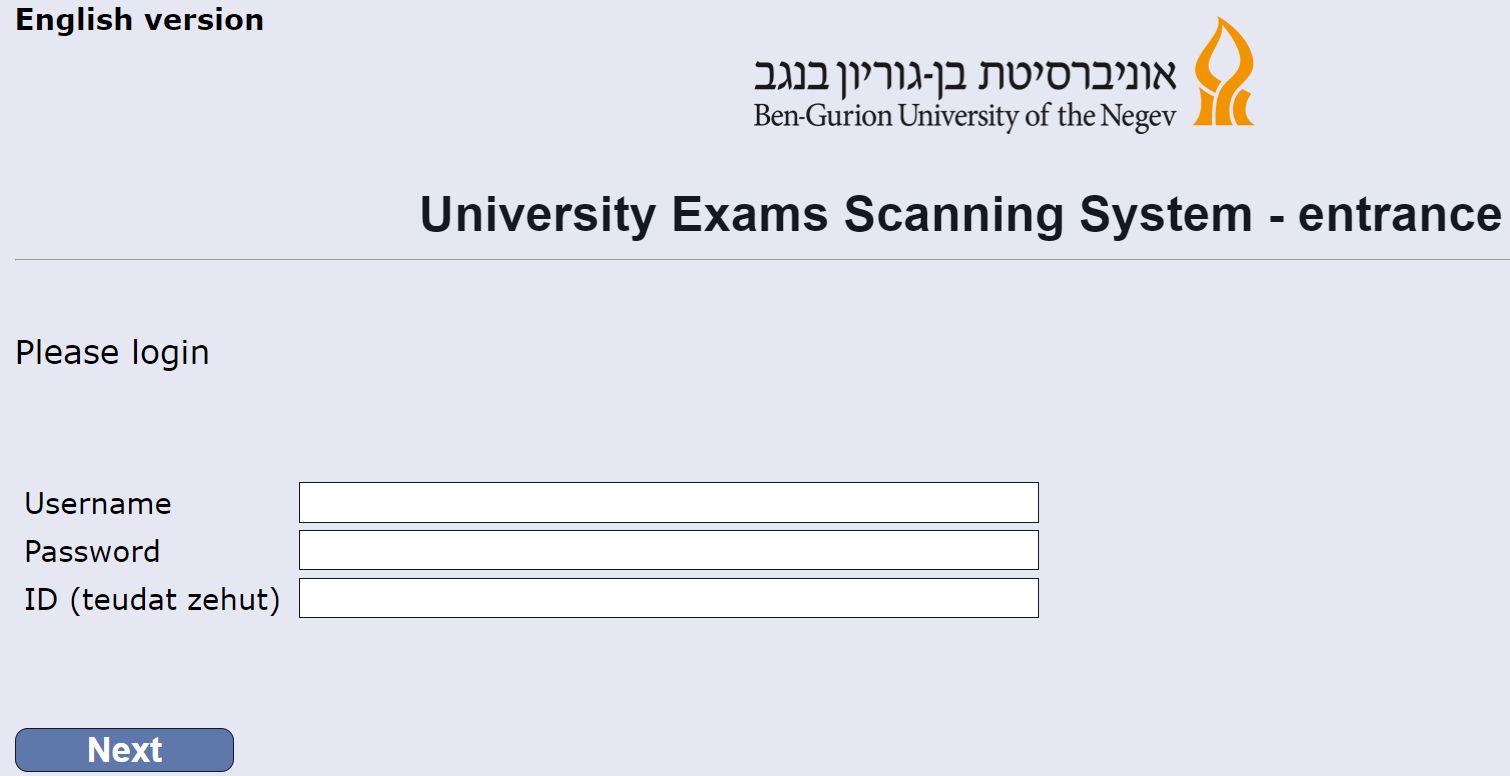}
        \caption{Phishing}
        \label{fig:phishing}
    \end{subfigure}
    \begin{subfigure}[t]{\textwidth}
    \centering
        \includegraphics[width=0.6\textwidth, keepaspectratio]{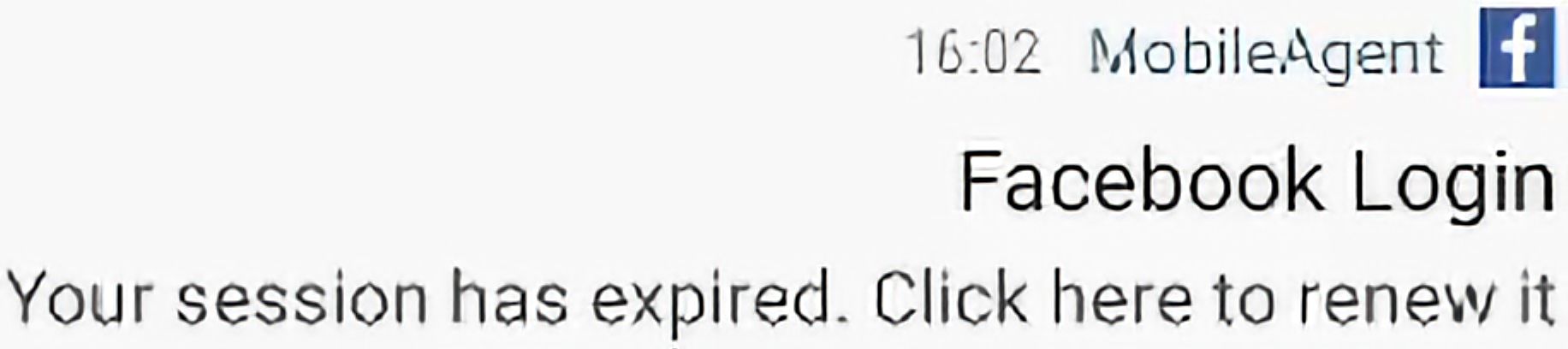}
        \caption{Impersonation}
        \label{fig:app_phish}
    \end{subfigure}
    
    \caption{Illustration of the different challenges} \label{fig:challenges}
\end{figure}

To evaluate \sysname's influence on behavior in active risk situations, we implemented three types of challenges: phishing, impersonation, and permission attacks.
The challenges were presented weekly, with one challenge of each type per week, resulting in three challenges every week and a total of 15 challenges. 
The order of the challenges presented during the week, as well as the day and hour in which they were presented, was randomized.
Examples of the challenges are provided in Figure~\ref{fig:challenges}. 
The challenges were designed as follows.

\noindent\emph{Phishing:}
Phishing is the most prevalent SE attack vector.
In our experiment, this attack involved creating a web page that emulates a login page from a pre-designed template, typically for student services. 
The attack was initiated by emailing the subjects and enticing them to click on an attached link to authenticate themselves for a supposed university-related event.
The link directed them to one of three domains that we purchased for the experiment, which resemble the actual university domain. 
The email was sent by a familiar sender, like `student administration,' who is known to the subjects as a legitimate email source for university administration. 
The email was sent during the academic semester when administrative emails from the university are expected.
Although the phishing email appears genuine, there are several indications that it was a phishing attack. 
First, it was not sent from the university's mail system; second, the link provided was not associated with the university's domain: and third, the phishing web page did not employ the HTTPS protocol. 
To safeguard the subjects' privacy, authentication information was not transmitted to the server. 
In this challenge, we evaluated the subject twice. 
First, we determined whether they clicked on the link and accessed the website. If they did, we then determined if they entered login details. The following phishing templates were used:

    \noindent\emph{(1) Facebook security}: An email was sent, informing subjects that they violated Facebook's code of conduct and their profile was at risk of deletion. 
    Subjects were urged to log in to their account and appeal, via a link provided in the email.
    
    \noindent\emph{(2) Moodle - new grade}: Moodle is a learning platform that the university uses to upload course materials and students use to submit assignments. 
    An email was sent to subjects telling them a grade was assigned to them on the Moodle platform, providing a link to log in and view it.
    
    \noindent\emph{(3) Organizational password change}: Students are required to change their organizational password periodically. 
    An email was sent to students asking them to change their password via the link provided or their account would be locked.
    
    \noindent\emph{(4) New appeal response}: In the subjects' university, students can make an appeal about the way in which their test was reviewed and graded. 
    During the exam period, an email was sent to subjects informing them of a response to an appeal they made regarding a specific exam, followed by a link to the appeal system.
    
    \noindent\emph{(5) New exam scan}: In the exam period, an email was sent to subjects telling them that an exam they took had been graded and the results were published.
    A link to the exam website was provided, enabling the subject to see the grade.
    
\noindent\emph{Permission attack:}
Malicious applications can trick unaware subjects into granting dangerous permissions during runtime. 
In each variant of this challenge, the device requested the granting of a dangerous permission to an app that does not need that permission. 
The mobile agent triggered the attack scenario when the subject used the phone and appeared on the screen using the Android permission requests' UI. 
The subject could reject or approve the request; a subject who granted privileges to the app was considered vulnerable to the attack.

The experiment included the following permission request templates: The Calculator requests camera permissions, WhatsApp requests calender permissions, Camera requests SMS permissions, and Gmail requests SMS permissions.

\noindent\emph{Impersonation:}
Fraudulent apps can deceive people in order to gain possession of their credentials.
In this challenge, we simulated a malicious application that sends a push notification while impersonating a legitimate service. 
The user interface of the notification exhibited a characteristic indicative of a phishing attack, which is the appearance of our mobile agent's name, along with the impersonated app name. 
Upon clicking the notification, our application launched, presenting a replica of the login screen of a well-known and trusted app.
To assess the subjects' performance in this attack, we classified them into two categories: half-vulnerable if they clicked the notification but did not complete the login process, and fully vulnerable if they both clicked the notification and completed the login process. 
To ensure the privacy of the subjects, the authentication information was not transmitted to the server.
The experiment included an app impersonating Facebook, Instagram, and the university's official app.

\subsection{Articles and Blog Posts}
\label{subsec:articles}
Prior to the experiment, we searched the web for publicly available relevant educational articles and blog posts.
We looked for two types of items: items about each focus area in the ISA taxonomy (meaning only about passive aspects) for the \sysname\space group and items about each of the three types of SE challenges (meaning only about active aspects) for the baseline group. After a thorough review, we found 32 items (16 per group) and labeled them by topic. 
Additionally, for the baseline group, each item was manually assigned a comprehensiveness grade, reflecting its depth and complexity on a scale from one (denoting basic and intuitive content) to five (indicating comprehensive and technical material).  
This grade determined the order in which the items were provided to subjects in the baseline group, as described in Section~\ref{sec:exp_setup}. 
In the \sysname\space group, the order of the items was predetermined and fixed for the entire training process. 
There was one item about each focus area in the ISA taxonomy. 
The list of articles and blog posts is presented in Table~\ref{table:articles} in the appendix.

\subsection{Experiment Setup}
\label{sec:exp_setup}
Each subject was assigned randomly to one of two groups, \sysname\space and baseline, each of which initially had 35 subjects. 
All subjects were asked to install our mobile app on their smartphones and keep it for the next five weeks. 
As mentioned in Section~\ref{chap:methodology}, we first needed to calculate an initial score for each subject in a calibration period.
All subsequent scores in the training process were relative to this initial score and used for personalization and later analysis.
For both groups, the calibration period consisted of the first week of the experiment. 
During this period, the mobile sensors monitored the subjects’ behavior, and they were presented with three challenges (one of each type). 
Afterward, the sensor monitoring and three weekly challenges continued until the end of the experiment. 
In addition, as mentioned in Section~\ref{subsec:challenges}, to compute the active score, we used a moving window of the last X challenges. 
We set X to be five for both groups. The training process began at the end of the calibration period and continued for four weeks.
Each group was trained using one of two different methods; a comparison of the groups' training processes is provided in Table~\ref{table:exp}.

\begin{table}
\caption{Comparison of the groups' training processes}
\label{table:exp}
\scriptsize
\centering
\begin{tabular}{|c|P{1.1cm}|P{1.1cm}|P{1.75cm}|p{3.2cm}|p{3.5cm}|}
\hline
\textbf{Group} &
\textbf{Subject of Articles} & 
\textbf{\# Articles} & 
\textbf{Gamification} & 
\textbf{Personalization} & 
\textbf{Timing} \\ \hline
\rule{0pt}{3ex} \multirow{1}{*}{\rotatebox[origin=c]{90} {\sysname}} & Passive risk related & 16 & \checkmark & 
The collection of articles is fixed. 
Low scores or negative score deltas direct subjects to articles related to focus areas that need improvement.
& All articles were available from the second week. \\
\hline
\multirow{1}{*}[-0.4cm]{\rotatebox[origin=c]{90} {Baseline}} & Active risk related & 8 are chosen personally, from a pool of 16 & \checkmark & 
Articles are selected based on the subject's performance in challenges. Their comprehensiveness increases with repeated failures in the same attack vector.
& Starting from the second week, articles were incrementally provided twice a week and remained accessible until the experiment's conclusion, with notifications sent to subjects' devices. \\ 
\hline
\end{tabular}
\end{table}

\subsection{Results}
\label{sec:results}
\begin{wrapfigure}{r}{0.5\textwidth}
    \centering
         \includegraphics[trim=0cm 1cm 0cm 6cm, width=0.5\textwidth,keepaspectratio]{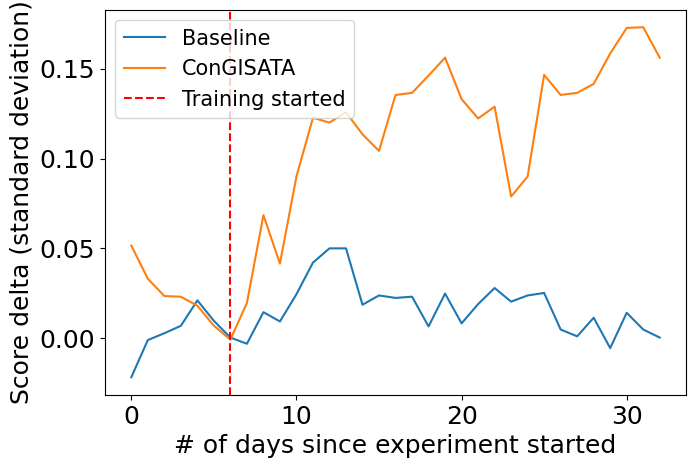}
          \captionsetup{belowskip=-25pt}
          \caption{Average passive score deltas per group over time}
         \label{fig:passive-score}
\end{wrapfigure}

In this study we address the following three research questions:

\textit{RQ1: Can our framework improve users' passive ISA score, as measured by the mobile ISA taxonomy? If so, how does it compare to the baseline method?}
First, we analyzed the passive score deltas and examined each criterion individually.
Figure~\ref{fig:sensors_result} (in the appendix) shows the delta in the score for each of the criteria as a function of the number of days since the experiment started. An increase in the score was observed for all but one criterion.
Furthermore, our framework resulted in a more notable improvement in the group's performance relative to that of the baseline group.
We also examined the total passive ISA score for each group, calculated as the average across the focus areas of the various criteria. As seen in Figure~\ref{fig:passive-score}, the use of our framework improved the passive ISA score, whereas no improvement was observed for the baseline group.

  \begin{wrapfigure}{r}{0.5\textwidth}
        \includegraphics[trim=0cm 1cm 0cm 2.6cm, width=0.5\textwidth]{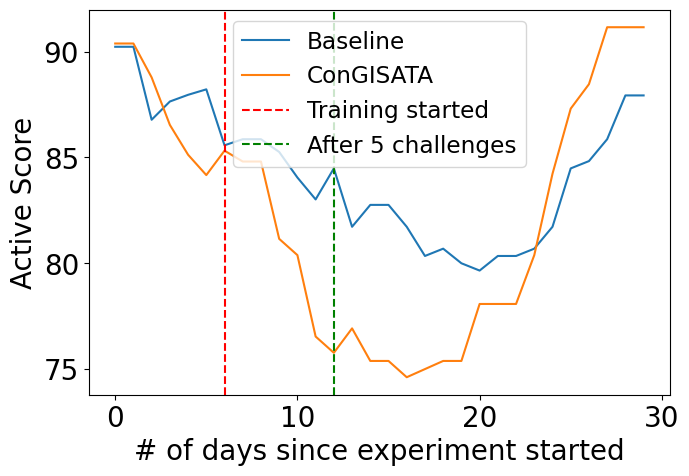}
        \captionsetup{belowskip=-25pt}
        \caption{Active score over time}
        \label{fig:active-score}
           \end{wrapfigure}
           
\textit{RQ2: Does \sysname\space help users improve their active ISA score, as measured using the challenges? If so, how does it compare to the baseline method?}
We analyzed the active score over time. 
Similar to other ISA training methods, the baseline method uses articles related to active risk situations thereby emphasizing active aspects. 
Thus, we anticipate that the active score of the baseline group will improve over time. 
Figure~\ref{fig:active-score} shows the change in score throughout the experiment. Initially, both groups experienced a decrease in their scores for two reasons: 
Firstly, during the first week (to the left of the red dotted line), the groups received no training. 
Secondly, the initial active score was calculated after day 13 (indicated by the green dotted line), after a sufficient number of challenges were presented -- a minimum of five challenges with at least one challenge from each one of the three attack vectors (see Section~\ref{sec:exp_setup}).
After day 13 both groups demonstrated notable improvement, with the \sysname\space group exhibiting slightly better performance. 
This result emphasizes that the training for secure passive behavior received by the \sysname\space group also reinforces active behavior.  

\begin{wrapfigure}{r}{0.5\textwidth}
    \centering
        \includegraphics[trim=0cm 1cm 0cm 0.3cm,width=0.5\textwidth,keepaspectratio]{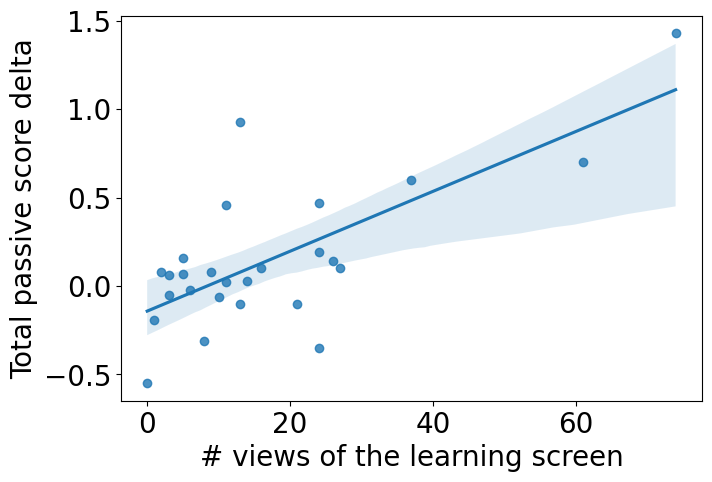}
        \captionsetup{belowskip=-22pt}
        \caption{Correlation between the number of views of the learning screen and passive score delta}
        \label{fig:app-score-corr}
\end{wrapfigure}

\textit{RQ3: Does increased use of our framework correlate with greater improvement in passive behavior?}

We logged every view of each of the app's screens and looked for a correlation between views and behavioral change. 
As expected, the most significant Pearson correlation was found between the number of views of the learning screen and the total delta in the passive score (r=0.72, p=3.41e-5), as seen in Figure~\ref{fig:app-score-corr}.
A similar result was obtained when checking for a correlation between the number of days in which a subject viewed the learning screen and the passive score delta. 
However, one of our learning screen's main advantages is its continuous nature, allowing users to see up-to-date details on each focus area with respect to their passive behavior. 
Going through the entire screen thoroughly may require more than one view per day so we chose to report the number of views and not the number of days, to differentiate subjects who viewed the learning screen multiple times a day from those who did not.

\section{Conclusion}

This study introduces \sysname\space - a continuous gamified ISA training and assessment framework that collects data from various sensors in users' everyday environments to examine aspects of ISA in real-life settings. 
The sensor readings are integrated into the framework, which generates a feedback loop. 
This continuous feedback mechanism helps users learn from their mistakes and improve their resilience against prevalent security risks. 
The use of sensors and challenges also provides a more reliable ISA assessment than the commonly used self-reported questionnaires. 
Our results confirm that \sysname\space improves passive behavior, while the baseline method does not. 
Moreover, although \sysname\space only provides articles on passive behavior, it helps users improve their ability to handle active attack scenarios. 
\sysname\space can be used in a corporate environment, in new employee training or as a regularly performed periodic procedure. Adapting the framework to new threats should be relatively easy, and may include these steps: (1) adding a new type of challenge simulating the new threat; (2) implementing additional sensors to measure related real-life behaviors; and (3) collecting (or creating) educational articles about the new threat. 
The number of subjects in this study does not allow meaningful analysis of the contribution of timing and personalization to \sysname's ability to improve ISA.  
This limitation can be addressed in more extensive experiments, including an ablation study performed with a large group of users, which we plan for future work.

\clearpage

\appendix
\section*{Appendix}
\subsection*{List of Articles and Blog Posts}
As described in Section~\ref{subsec:articles}, we collected 32 publicly available relevant educational articles and blog posts to use in the experiment (the blog posts and articles are listed in Table~\ref{table:articles}). The items for the ConGISATA group are listed first, with their corresponding ISA taxonomy criterion ID, and do not include a comprehensiveness grade. The items for the baseline group, which include a comprehensiveness grade, are listed after the bold horizontal line.
\begin{table}
\caption{The articles and blog posts used in the experiment}
\label{table:articles}
\begin{tabular}{|p{0.5cm}|p{4.5cm}|p{3.5cm}|p{3.3cm}|}
\hline
& \textbf{Topic} & \textbf{Links} & \textbf{Comprehensiveness Grade}\\
\hline
\multirow{16}{*}{\rotatebox[origin=c]{90} {ConGISATA}}  &
Account (A2) & \href{https://www.bu.edu/tech/support/information-security/why-use-2fa/}{link} & - \\
\cline{2-4}
& Account (A3) & \href{https://cybernews.com/best-password-managers/how-do-password-managers-work/}{link} & - \\
\cline{2-4}
& Browser (B1) & \href{https://www.mimecast.com/blog/what-are-malicious-websites/}{link} & - \\
\cline{2-4}
& Virtual Communication (VC1) & \href{https://www.givaudan.com/specials/infosec/tip-05}{link} & - \\
\cline{2-4}
& Virtual Communication (VC2) & \href{https://www.phishing.org/what-is-phishing}{link} & - \\
\cline{2-4}
& Network (N1) & \href{https://www.givaudan.com/specials/infosec/tip-02}{link} & - \\
\cline{2-4}
& Network (N3) & \href{https://www.kaspersky.com/resource-center/threats/why-use-vpn-on-smartphone}{link} & - \\
\cline{2-4}
& Application Installation (AI1) & \href{https://www.makeuseof.com/tag/safe-install-android-apps-unknown-sources/}{link} & - \\
\cline{2-4}
& Application Installation (AI2) & \href{https://www.avg.com/en/signal/guide-to-android-app-permissions-how-to-use-them-smartly}{link} & - \\
\cline{2-4}
& Application Installation (AI3) & \href{https://tapadoo.com/mobile-app-ratings-reviews/}{link} & - \\
\cline{2-4}
& Application Handling (AH1) & \href{https://www.getcybersafe.gc.ca/en/blogs/software-updates-why-they-matter-cyber-security}{link} & - \\
\cline{2-4}
& Application Handling (AH3) & \href{https://www.howtogeek.com/778951/why-you-should-get-rid-of-unused-android-apps/}{link} & - \\
\cline{2-4}
& Security Systems (SS2) & \href{https://www.ncsc.gov.uk/guidance/what-is-an-antivirus-product}{link} & - \\
\cline{2-4}
& Security Systems (SS5) & \href{https://www.totaldefense.com/security-blog/the-importance-of-having-a-lock-screen-on-your-device/}{link} & - \\
\cline{2-4}
& Physical Connectivity (PC1) & \href{https://www.wired.com/story/turn-off-bluetooth-security/}{link} & - \\
\cline{2-4}
& Operating System (OS2) & \href{https://www.bullguard.com/bullguard-security-center/mobile-security/mobile-threats/android-rooting-risks.aspx}{link} & - \\
\Xhline{5\arrayrulewidth} \multirow{7}{*}{\rotatebox[origin=c]{90} {Baseline}}
& Impersonation Attacks & \href{https://www.givaudan.com/specials/infosec/tip-03}{link} & 2 \\ 
\cline{2-4}
& Impersonation Attacks & 
                        \href{https://www.reshiftmedia.com/avoid-phishing-scammers-impersonating-facebook/}{link},
                        \href{https://www.mcafee.com/blogs/privacy-identity-protection/how-to-spot-fake-login-pages/}{link},
                        \href{https://mnlb.bank/steer-clear-of-fake-login-pages/}{link},
                        \href{https://blog.icorps.com/grayware-app-safety-draft}{link} & 3 \\
\cline{2-4}
& Impersonation Attacks & \href{https://heimdalsecurity.com/blog/malicious-app-definition/}{link} & 5 \\
\cline{2-4}
& Permission Attacks & \href{https://www.avg.com/en/signal/guide-to-android-app-permissions-how-to-use-them-smartly}{link},
                     \href{https://www.comparitech.com/blog/vpn-privacy/secure-android-app-permissions/}{link} & 2 \\
\cline{2-4}
& Permission Attacks & \href{https://www.trendmicro.com/vinfo/pl/security/news/mobile-safety/12-Most-Abused-Android-App-Permissions}{link},
                     \href{https://blog.f-secure.com/3-sketchy-app-permissions-and-how-to-stop-them-from-ruining-your-day/}{link} & 3 \\
\cline{2-4}
& Permission Attacks & \href{https://blog.nviso.eu/2021/09/01/how-malicious-applications-abuse-android-permissions/}{link} & 5 \\
\cline{2-4}
& Phishing Attacks & \href{https://social-sciences.tau.ac.il/PAYPAL}{link},
                   \href{https://social-sciences.tau.ac.il/Vishing}{link},
                   \href{https://social-sciences.tau.ac.il/fishing}{link},
                   \href{https://social-sciences.tau.ac.il/phishing}{link},
                   \href{https://social-sciences.tau.ac.il/SMISHING}{link} & 1 \\
\hline

\end{tabular}
\end{table}

\subsection*{Passive Score Delta by Criterion}
\begin{figure}
\includegraphics[height=0.9\textheight, keepaspectratio]{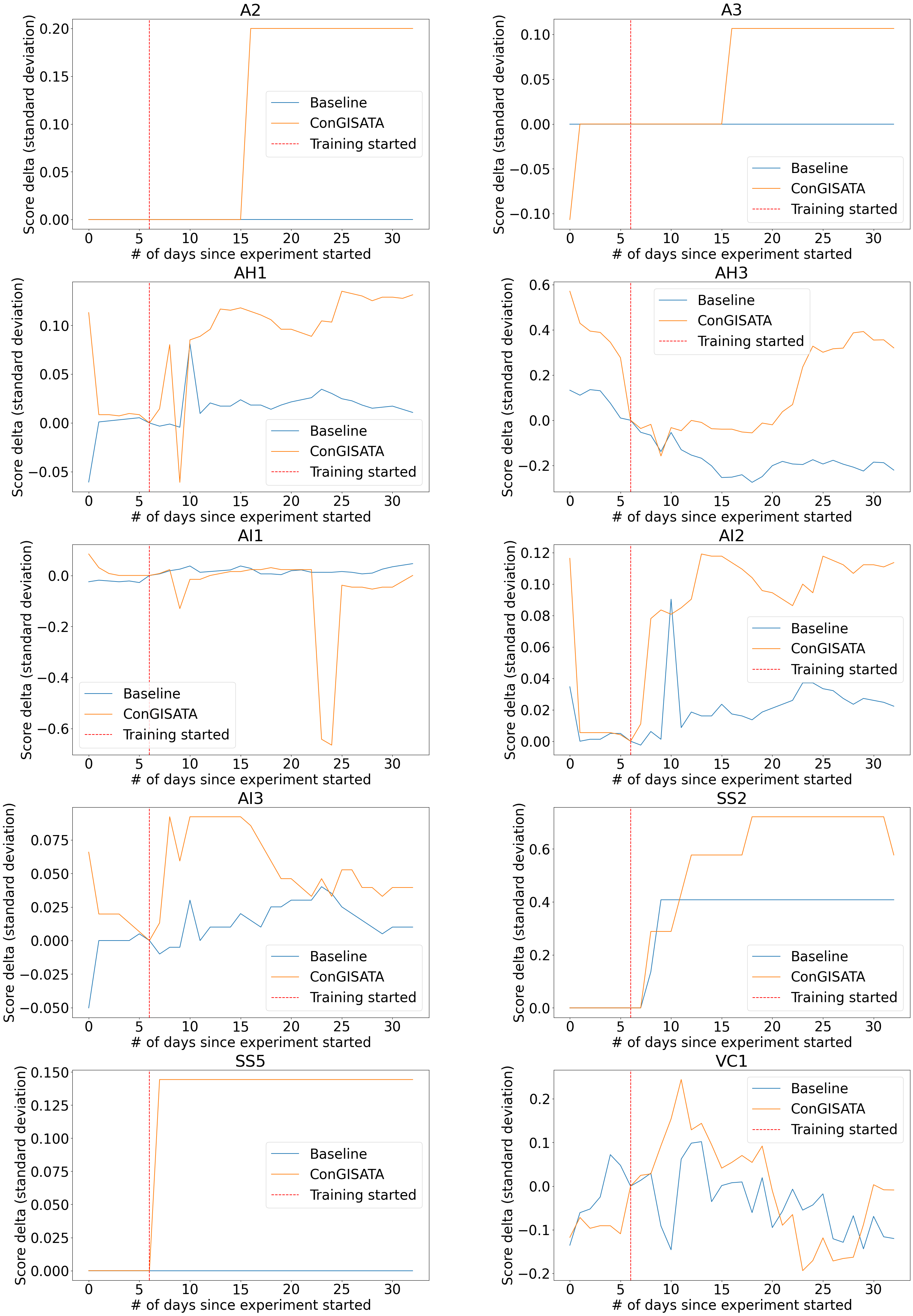}
       \caption{Average score deltas for the groups per criterion, as a function of the number of days since the experiment started.}
        \label{fig:sensors_result}
\end{figure}

Figure~\ref{fig:sensors_result} shows the average score deltas for the groups per criterion, as a function of the number of days since the experiment started.

\clearpage
\begin{thebibliography}{33}
\bibitem{kumar2015social}Kumar, A., Chaudhary, M. \& Kumar, N. Social engineering threats and awareness: a survey. {\em European Journal Of Advances In Engineering And Technology}. 2, 15-19 (2015)
\bibitem{kelly2017almost}Kelly, R. Almost 90\% of cyber attacks are caused by human error or behavior. {\em ChiefExecutive. Net}. (2017)
\bibitem{bada2019cyber}Bada, M., Sasse, A. \& Nurse, J. Cyber security awareness campaigns: Why do they fail to change behaviour?. {\em ArXiv Preprint ArXiv:1901.02672}. (2019)
\bibitem{deterding2011game}Deterding, S., Dixon, D., Khaled, R. \& Nacke, L. From game design elements to gamefulness: defining" gamification". {\em Proceedings Of The 15th International Academic MindTrek Conference: Envisioning Future Media Environments}. pp. 9-15 (2011)
\bibitem{hamari2014does}Hamari, J., Koivisto, J. \& Sarsa, H. Does gamification work?–a literature review of empirical studies on gamification. {\em 2014 47th Hawaii International Conference On System Sciences}. pp. 3025-3034 (2014)
\bibitem{gjertsen2017gamification}Gjertsen, E., Gjære, E., Bartnes, M. \& Flores, W. Gamification of Information Security Awareness and Training.. {\em ICISSP}. pp. 59-70 (2017)
\bibitem{kumaraguru2009school}Kumaraguru, P., Cranshaw, J., Acquisti, A., Cranor, L., Hong, J., Blair, M. \& Pham, T. School of phish: a real-world evaluation of anti-phishing training. {\em Proceedings Of The 5th Symposium On Usable Privacy And Security}. pp. 1-12 (2009)
\bibitem{bitton2018taxonomy}Bitton, R., Finkelshtein, A., Sidi, L., Puzis, R., Rokach, L. \& Shabtai, A. Taxonomy of mobile users' security awareness. {\em Computers \& Security}. 73 pp. 266-293 (2018)
\bibitem{keinan2012leaving}Keinan, R. \& Bereby-Meyer, Y. " Leaving it to chance"–Passive risk taking in everyday life.. {\em Judgment \& Decision Making}. 7 (2012)
\bibitem{keinan2017perceptions}Keinan, R. \& Bereby-Meyer, Y. Perceptions of active versus passive risks, and the effect of personal responsibility. {\em Personality And Social Psychology Bulletin}. 43, 999-1007 (2017)
\bibitem{bitton2020evaluating}Bitton, R., Boymgold, K., Puzis, R. \& Shabtai, A. Evaluating the Information Security Awareness of Smartphone Users. {\em Proceedings Of The 2020 CHI Conference On Human Factors In Computing Systems}. pp. 1-13 (2020)
\bibitem{newbould2009playing}Newbould, M. \& Furnell, S. Playing Safe: A prototype game for raising awareness of social engineering. {\em Australian Information Security Management Conference}. pp. 4 (2009)
\bibitem{hart2020riskio}Hart, S., Margheri, A., Paci, F. \& Sassone, V. Riskio: A Serious Game for Cyber Security Awareness and Education. {\em Computers \& Security}. pp. 101827 (2020)
\bibitem{chapman2014picoctf}Chapman, P., Burket, J. \& Brumley, D. PicoCTF: A game-based computer security competition for high school students. {\em 2014 {USENIX} Summit On Gaming, Games, And Gamification In Security Education (3GSE 14)}. (2014)
\bibitem{denning2013control}Denning, T., Lerner, A., Shostack, A. \& Kohno, T. Control-Alt-Hack: the design and evaluation of a card game for computer security awareness and education. {\em Proceedings Of The 2013 ACM SIGSAC Conference On Computer \& Communications Security}. pp. 915-928 (2013)
\bibitem{alqahtani2020design}Alqahtani, H. \& Kavakli-Thorne, M. Design and Evaluation of an Augmented Reality Game for Cybersecurity Awareness (CybAR). {\em Information}. 11, 121 (2020)
\bibitem{luh2020penquest}Luh, R., Temper, M., Tjoa, S., Schrittwieser, S. \& Janicke, H. PenQuest: a gamified attacker/defender meta model for cyber security assessment and education. {\em Journal Of Computer Virology And Hacking Techniques}. 16, 19-61 (2020)
\bibitem{yasin2018improving}Yasin, A., Liu, L., Li, T., Fatima, R. \& Jianmin, W. Improving software security awareness using a serious game. {\em IET Software}. 13, 159-169 (2018)
\bibitem{arend2020passive}Arend, I., Shabtai, A., Idan, T., Keinan, R. \& Bereby-Meyer, Y. Passive-and Not Active-Risk Tendencies Predict Cyber Security Behavior. {\em Computers \& Security}. pp. 101929 (2020)
\bibitem{selvam2020human}Selvam, V. Human Error in IT Security. {\em ArXiv Preprint ArXiv:2005.04163}. (2020)
\bibitem{dunlosky2013improving}Dunlosky, J., Rawson, K., Marsh, E., Nathan, M. \& Willingham, D. Improving students’ learning with effective learning techniques: Promising directions from cognitive and educational psychology. {\em Psychological Science In The Public Interest}. 14, 4-58 (2013)
\bibitem{canham2022phish}Canham, M., Posey, C. \& Constantino, M. Phish Derby: Shoring the Human Shield Through Gamified Phishing Attacks. {\em Frontiers In Education}. 6 pp. 536 (2022)
\bibitem{jaffray2021sherlocked}Jaffray, A., Finn, C. \& Nurse, J. SherLOCKED: A Detective-Themed Serious Game for Cyber Security Education. {\em International Symposium On Human Aspects Of Information Security And Assurance}. pp. 35-45 (2021)
\bibitem{Threat-Report-2023}Sophos Sophos 2023 Threat Report.  (2022), https://assets.sophos.com/X24WTUEQ/at/b5n9ntjqmbkb8fg5rn25g4fc/sophos-2023-threat-report.pdf
\bibitem{redmiles2018asking}Redmiles, E., Zhu, Z., Kross, S., Kuchhal, D., Dumitras, T. \& Mazurek, M. Asking for a friend: Evaluating response biases in security user studies. {\em Proceedings Of The 2018 Acm Sigsac Conference On Computer And Communications Security}. pp. 1238-1255 (2018)
\bibitem{solomon2022contextual}Solomon, A., Michaelshvili, M., Bitton, R., Shapira, B., Rokach, L., Puzis, R. \& Shabtai, A. Contextual security awareness: A context-based approach for assessing the security awareness of users. {\em Knowledge-Based Systems}. 246 pp. 108709 (2022)
\bibitem{bockle2017towards}Böckle, M., Novak, J. \& Bick, M. Towards adaptive gamification: a synthesis of current developments.  (2017)
\bibitem{alahmari2022moving}Alahmari, S., Renaud, K. \& Omoronyia, I. Moving beyond cyber security awareness and training to engendering security knowledge sharing. {\em Information Systems And E-Business Management}. pp. 1-36 (2022)
\bibitem{dincelli2020choose}Dincelli, E. \& Chengalur-Smith, I. Choose your own training adventure: designing a gamified SETA artefact for improving information security and privacy through interactive storytelling. {\em European Journal Of Information Systems}. 29, 669-687 (2020)
\bibitem{scholefield2019gamification}Scholefield, S. \& Shepherd, L. Gamification techniques for raising cyber security awareness. {\em HCI For Cybersecurity, Privacy And Trust: First International Conference, HCI-CPT 2019, Held As Part Of The 21st HCI International Conference, HCII 2019, Orlando, FL, USA, July 26–31, 2019, Proceedings 21}. pp. 191-203 (2019)
\bibitem{omar2021malware}Omar, N., Foozy, C., Hamid, I., Hafit, H., Arbain, A. \& Shamala, P. Malware awareness tool for internet safety using gamification techniques. {\em Journal Of Physics: Conference Series}. 1874 pp. 012023 (2021)
\bibitem{wu2021assessing}Wu, T., Tien, K., Hsu, W. \& Wen, F. Assessing the effects of gamification on enhancing information security awareness knowledge. {\em Applied Sciences}. 11, 9266 (2021)
\bibitem{heid2020raising}Heid, K., Heider, J. \& Qasempour, K. Raising Security Awareness on Mobile Systems through Gamification. {\em Proceedings Of The European Interdisciplinary Cybersecurity Conference}. pp. 1-6 (2020)

\end{thebibliography}
\end{document}